\documentclass{raa}
\usepackage{graphicx,times}             
\input{epsf.sty}                        

\begin{document}

   \title{DETERMINATION OF THE CHROMOSPHERIC QUIET NETWORK ELEMENT AREA INDEX AND ITS 
VARIATION DURING 2008-2011}
  \volnopage{Vol.0 (200x) No.0, 000--000}      
   \setcounter{page}{1}          

   \author{Jagdev Singh
      \inst{1}
   \and B. Ravindra
   \and R. Selvendran
    \and P. Kumaravel
    \and M. Priyal
    \and T. G. Priya
    \and K. Amareswari
   }

   \institute{Indian Institute of Astrophysics, Koramangala, Bengaluru-560 034, INDIA {\it jsingh@iiap.res.in} \\
}
   \date{Received~~2011 August; accepted~~2011~~month day} 

\abstract{Generally it has been considered that the plages and sunspots are the main contributors 
to the solar irradiance. There are small scale structures on the sun with intermediate magnetic fields
that could also contribute to the solar irradiance. It has not yet been quantified how much of these 
small scale 
structures contribute to the solar irradiance and how much it varies over the solar cycle.
In this paper, we used Ca~II~K images obtained from the telescope installed at Kodaikanal observatory.
We report a method to separate the network elements from the background structure 
and plage regions.  We compute the changes in the network element area index during the minimum
phase of solar cycle and part of the ascending phase of cycle 24. The measured 
area occupied by 
the network elements is about 30\% and plages less than 1\% of the solar disk during the observation 
period from February 2008 -2011. During the extended period of minimum activity it is observed that the network element area index decreases by about 7\% compared to the area occupied by the network elements 
in 2008. A long term study of network element area index is required to understand the variations over 
the solar cycle.}


   \authorrunning{Singh et~al. }            
   \titlerunning{variations in network element area index}  

   \maketitle

\section{Introduction}           
\label{sect:intro}
The synoptic observations of the sun is very important to study the long term
variations of the solar magnetic field and its effect on space weather and climate.
The large and small scale structures observed on the sun contributes to the solar 
irradiances in different magnitudes at different times of the solar cycle 
(Worden, White and Woods,~\cite{Worden+White+Woods1998}). 
The small scale network structures can be observed at the chromospheric level and above
in the solar atmosphere. The contrast of the network structures is large in Ca~II~K wavelength 
band. The main features we observe in the line center of Ca-K images are the plages, active network,
enhanced network and network regions. Many have used the Ca-K images to estimate the contribution 
of the plages and network to the total solar irradiance at different times of the 
epoch (Worden, White and Woods,~\cite{Worden+White+Woods1998}).  Several authors have
used different techniques to compute the Ca-K plage index from the data of
Kodaikanal, Mt. Wilson, Arcetri and Sac Peak observatories and compared the
results obtained from different data sets 
(Foukal et~al.,~\cite{Foukaletal2009}, Tlatov, Pevtsov, and Singh,~\cite{Tlatov+Pevtsov+Singh2009}, Ermolli et~al.,~\cite{Ermollietal2009}).  

Many techniques have been developed to estimate the area of the network. Most of these 
results are restricted to a small time period of data.
Hagenaar, Schrijver, and Title~(\cite{hagenaar97}) used
a basin finding algorithm to estimate the areas of supergranular network cells. The application
of this technique on the Ca-K images gave a cell size of about 15~Mm. Similar technique used by 
Srikanth, Singh, and Raju~(\cite{srikanth2000}) showed that 
a cell size of 25~Mm, close to the 
most commonly known size of the network cells. Schrijver, Hagenaar, and Title~(\cite{schrijver97}) used
a gradient based tessellation algorithm to the Ca-K images to find the network boundaries. This
is very effective technique to find the area of the network cells. The skeleton 
method (Berrilli, Florio, and Ermolli,~\cite{berrilli1998}) along with a cellular automaton method 
(Berrilli et~al.,~\cite{berrilli1999} )
applied to the Precision Solar Photometric Telescope (PSPT) Ca-K high pass filtered binary images showed
that the area distribution of the network cells. From this method it is found that the network 
cells have a characteristic cell sizes of about 24~Mm.  

The techniques 
include the gradient based tessellation algorithms (Schrijver, Hagenaar, and Title, \cite{schrijver97}), image decomposition 
method (Harvey and White,~\cite{harvey99}, Worden, White, and Woods,~\cite{Worden+White+Woods1998}), skeletonizing iterative procedure (Ermolli, Berrilli, and Florio,~\cite{ermolli2003}),
to measure the plage and network areas are restricted to small time intervals 
and do not include one complete solar cycle. However, many others 
(Tlatov, Pevtsov, and Singh,~\cite{Tlatov+Pevtsov+Singh2009} )
have tried to 
estimate the variation in the plage indices over several solar cycles using the digitized data sets 
of Kodaikanal, Mt. Wilson etc. Though the overall pattern in the variation of the plage areas 
match well from each of the observatories, their magnitudes seems to be different for different 
observatory data sets (Ermolli, Solanki, and Tlatov,~\cite{Ermollietal2009}).

From the measurements on 2.4$^{\prime\prime}$ square aperture, Skumanich, Smythe and Frazier 
(\cite{Skumanich+Smythe+Frazier1975})
showed that the average network has brightness of 1.27 to that of the average non-network 
undisturbed chromosphere and cover 39\% of the quiet Sun. They probably refer 
to the active network. Worden, White, and Woods~(\cite{Worden+White+Woods1998}) have analyzed 1400 Ca-K line 
spectroheliograms covering parts of solar cycle 21 and 22 with a spatial resolution
8.5$^{\prime\prime}$. They define structures in Ca-K line spectroheliograms in 4 categories, 
plages, enhanced network, active network and quiet sun. They systematically identify 
these components using different intensity threshold and the minimum size for each 
category. The identified plages show intensity contrast relative to the quiet-sun 
greater than 2.776 and minimum value of 1.44. Similarly, the upper and lower level intensity 
contrasts are 2.06 and 1.37 for the enhanced network and 1.59 and 1.18 for the active 
network. They have built up the masks of plages and found the plage area to be larger 
by a factor of 2.8 as compared to the plage areas published by NGDC.

Most of the time enhanced or active network is present on the Ca-K line images 
and this makes it difficult to separate network elements. 
The Kodaikanal has a history of observing the full-disk images of the sun since  1907.
Recently, it has been updated with a filter based images with charge couple device as 
a image acquisition system. The filter centered at 393.37 nm of the Ca-K wavelength 
has been used to image the sun. The data obtained with this telescope provide a good 
opportunity to estimate the network element area index and its variation with time during the minimum 
solar activity period. We used a global fit method to Ca-K data to normalize the images and 
used the threshold limits to identify the network elements and plages on the
solar surface automatically.  In this case, because of extended solar minimum during 2008-2010, 
we have only the quite network features and small amount of plages with out the presence 
of active and enhanced network. Hence, it is a good opportunity to study the network element area
index. In this paper, we give a brief description about the instrument used to obtain the Ca-K 
images at Kodaikanal observatory. We also present a result of the Ca-K network element area 
index variations during the period between February 2008 and February 2011 
spanning over about 3 years. In the next Section, 
we describe our instrument used for obtaining the Ca-K data sets at Kodaikanal 
observatory. The results on variation of 
the network element area index is presented in Section 3 and in Section 4 we discuss the 
various other studies and compare those studies with the result obtained by us. We summarize the 
results in the last section.

\section{Instrument and Observations}
At Kodaikanal observatory we have been obtaining on daily basis Photohelio-grams 
of the sun since 1904 using a 15-cm aperture telescope, Ca-K spectroheliograms 
since 1907 and H$_{\alpha}$ spectroheliograms since 1912. Earlier, the images of the 
sun were recorded on specialized photographic emulsion. In 1995 we started using at Kodaikanal, 
a narrow band filter but using the old siderostat and CCD camera of 1K$\times$1K format to take 
Ca-K filtergrams. We now designed and fabricated a
telescope to take Ca-K line and white light images of the sun and named it as TWIN
telescope. This telescope has been in operation since 2008 at our Kodaikanal
observatory and collecting images during the clear skies.

The Twin telescope consists of 2 telescopes mounted on a single tracker. One of the
telescope takes the white light images in the continuum  of the blue wavelength and the 
other one is observing the sun in Ca-K wavelength. The focal length of the 150~mm objective 
lens from Zeiss is 2250~mm that forms the solar image of 20.6~mm size. The optical system are 
capable of providing a spatial resolution of 0.7 arcsec at the wavelength of 400~nm. A interference 
filter with 10~nm pass band centered at 395nm is kept in-front of the objective lens to avoid the excess 
heat in the telescope that also reduce the intensity to the required level to record the solar images. 
A thermally controlled narrow band interference filter centered at 393.37 nm with a 
passband of 0.12 nm is kept near the focus to isolate central portion of the Ca-K line. The CCD 
camera with a 2K$\times$2K pixel format of 16-bit read out at 1 MHz yield the images in Ca-K.
The image scale is 93$^{\prime\prime}$ per mm and thus pixel size with 13.5~$\mu$m provides a 
resolution of 1.25~arcsec per pixel. The images have been taken at a rate of 1 in 5 minutes.
In case of activity on the sun such as flares, we plan to obtain the images
at an interval of 2 seconds by restricting the FOV and binning the CCD chip by
2$\times$2 pixel.

The Twin telescope was made functional since February 23, 2008 and is obtaining the Ca-K line images 
of the sun whenever the sky is clear. We have analyzed the data till February 28, 2011. 
The quality of the data is mostly homogeneous except a couple of gaps when the observations 
could not be made. The data obtained in cloudy conditions have 
not been used in this study. On a full clear day we were able to get about 80 images. 
 In all we have used about 16000 images in our study. We compute the average network 
element area for the day using all the images taken on that day.  The obtained data are dark 
subtracted and corrected for flatfielding.

\section{Results}   

\begin{figure}
\begin{center}
\includegraphics[width=0.5\textwidth]{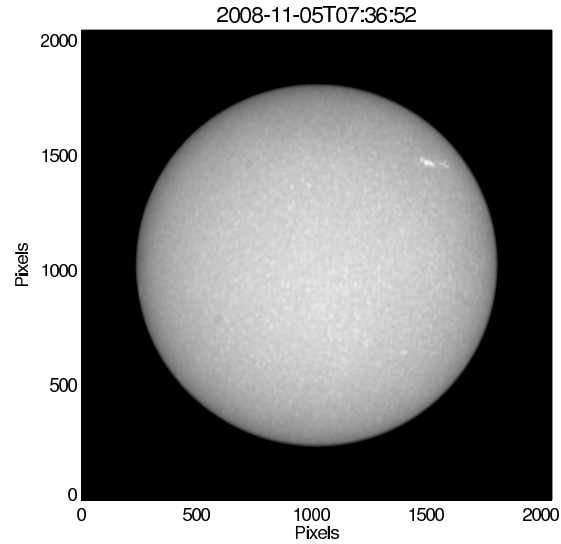}
\end{center}
\caption{Context image of Ca-K taken from the Twin telescope is shown here for the date of observation 
of 2008-November-05.}
\label{fig:1}
\end{figure}

A sample of the observed Ca-K image (Figure \ref{fig:1}) shows network structure, a small 
plage region near the limb and intensity gradient due to limb darkening effect.
No other type of large spatial scale intensity gradients are seen. When we started the 
solar observations with this telescope, the solar activity was 
in its declining phase. Hence, we do not see many active regions on the solar disk.

\subsection{Identification of network cells}
In order to identify the network elements we followed the procedure as below: (1) limb
fitting to identify the image center and radius of the solar disk (2) computing the 
limb darkening profiles (3) fitting a polynomial curve to the intensity profile 
and (4) thresholding method to identify the network cells and plages.
The details of this method are described below:

\subsubsection{Limb fitting method to identify the disk center and radius}
In the technique of automatically detecting the features on the sun, it is 
first essential to identify the center and radius of the sun in terms of pixels.
Hence, we first computed the radius and center of the solar disk in the calibrated data sets. 
This has been achieved by identifying the solar limb. The solar limb has a steep gradient 
between the solar disk and the surrounding region. We used the sobel filter to detect 
the edge of the solar limb.  
\begin{figure*}
\begin{center}
\includegraphics[width=0.45\textwidth]{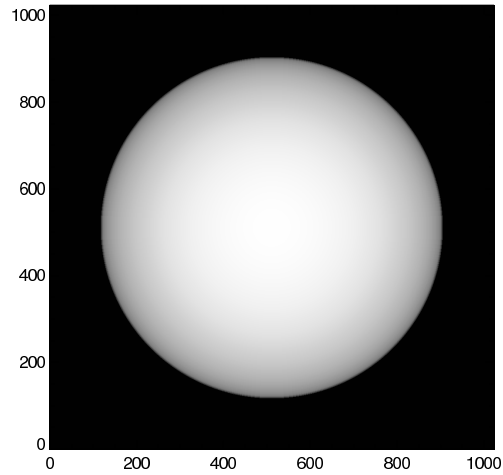} \\
\end{center}
\caption{The computed two-dimensional limb darkening function used to remove the 
limb darkening effect in Ca-K image.}
\label{fig:2}
\end{figure*}
We then used a threshold value of 5 times the mean value of the sobel filtered image to retain only the 
edges of the solar limb. We then detected the 8 points on the limb, 4 points on East, West,
North and South each and other 4 points are in  S-E, N-E, N-W and S-W points. Using the 8 
detected points on the solar limb we used a polynomial fitting to 
obtain the sun center and radius in terms of pixels. With this 
method we could identify the sun center and solar limb with one pixel accuracy.

\subsubsection{Removal of limb darkening}

The intensity distribution is not uniform over the sun's disk. It is large in the disk
center and decreases towards the limb. In order to remove the non uniformity in the
intensity due to the limb darkening effect we need to compute the limb darkening profile. 
This has been achieved by the method described in Denker et al.~(\cite{Denker99}). 
Those procedures have been adopted 
to remove the limb darkening in the H$_\alpha$ images. A similar procedure has been 
adopted here. We have used the median value of intensity
and it is less sensitive to asymmetric intensity distribution. To do this, the 
full-disk images are transformed to polar coordinates and at each radial position 
the median value is computed. This way it gives the average radial profile and it is smoothed 
by large kernels. The final averaged
and smoothed profile has been transformed to Cartesian coordinates. A resulting limb 
darkening image is shown in Figure \ref{fig:2}. We have also shown the profile of the
limb darkening corrected image in Figure \ref{fig:3}(left). This profile has been extracted
for those pixels shown as dark line in the middle of Figure \ref{fig:3}(right). The profile in 
Figure \ref{fig:3}(left) shows that in addition to small amplitude due to small scale 
variations in the network, a small amount of residual gradient is present. We remove these 
gradient by fitting a polynomial as described in the next section. The limb darkening 
function is computed for each image. The correction is applied to the calibrated data set to 
remove the limb darkening in the images. 

\subsubsection{Polynomial fitting and thresholding}
\label{pft}

\begin{figure*}
\begin{center}
\includegraphics[width=0.45\textwidth]{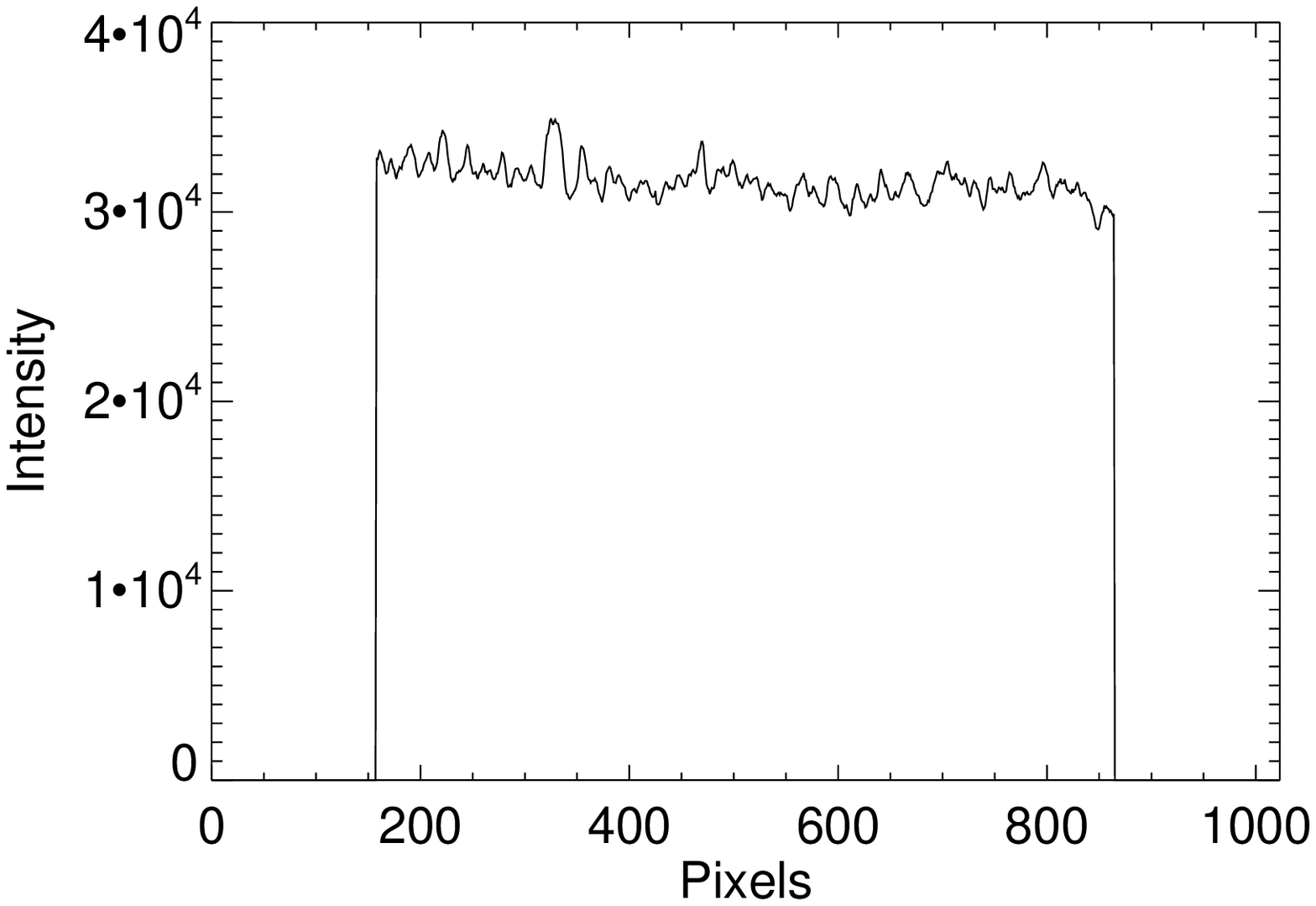}\includegraphics[width=0.6\textwidth]{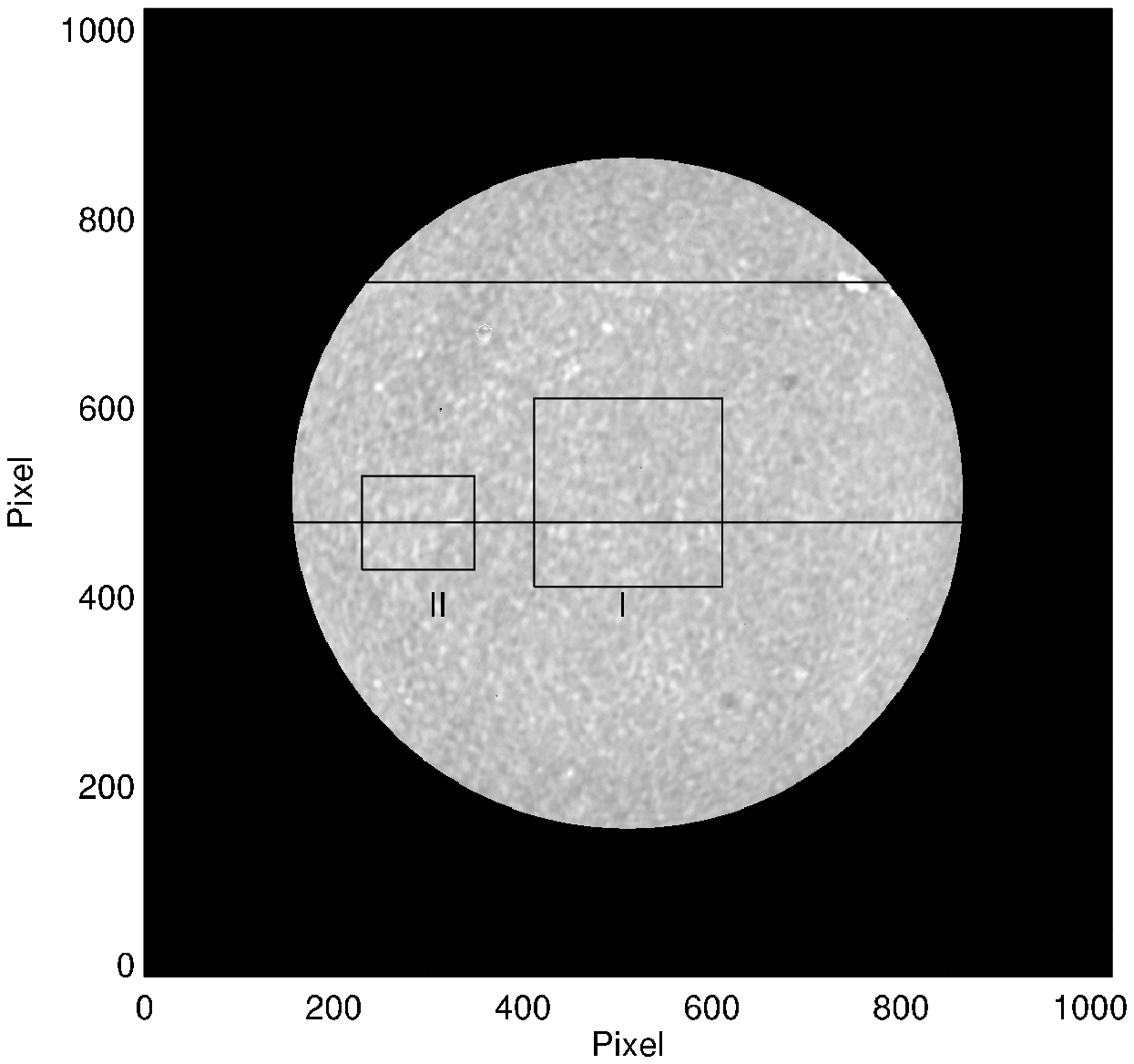} \\
\end{center}
\caption{Left: A typical intensity profile along a row of the image corrected for 
limb-darkening effect. Right: The image obtained after the limb darkening correction
and polynomial fitting. The boxes in black color marked on the  
Ca-K image represent the regions for which the contours of 
network elements are shown in subsequent Figures. The black lines are the 
locations for which the profiles has been shown in Figure \ref{fig:4}.}
\label{fig:3}
\end{figure*}

The solar image in Ca-K displays several features such as network elements, intra-network region 
and plages.
Network cells have borders and they are intense than the intra-network points.
However, the network elements are less in brightness than 
the plage regions in the images taken in Ca-K filters.
We used only the 90\% of the inner solar disk in the
network element/plage area index estimation as the fitted profile 
near the edges of the image leads to large uncertainty due to sharp intensity gradient.
Now to segregate each of the features in solar image we have used the
solar images after making corrections for the limb darkening as described above.  
Further, In order to remove the residual 
intensity gradient in the images we repeatedly applied a 3rd degree polynomial to 1-dimensional
row wise as well as column wise sliced pixels by keeping the mean level of the 1-D data as 
before. By  this method we make the background uniform and there are no variations in the
background image. This procedure keeps only a small scale pixel to pixel intensity 
variations due to the network elements and plages only. The resulting 
image is shown in Figure \ref{fig:3}(right).  

\begin{figure*}
\begin{center}
\includegraphics[width=0.4\textwidth]{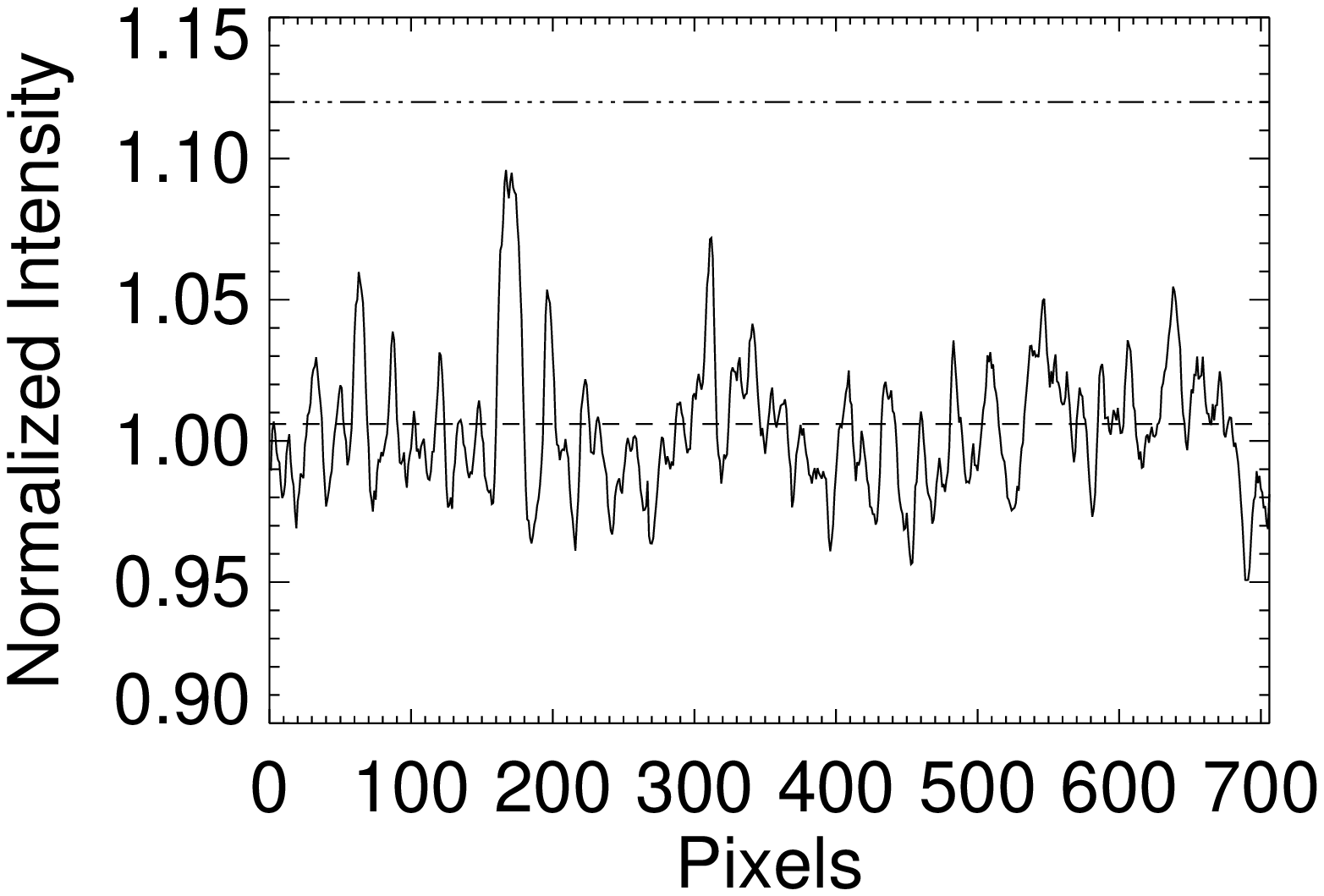}\includegraphics[width=0.37\textwidth]{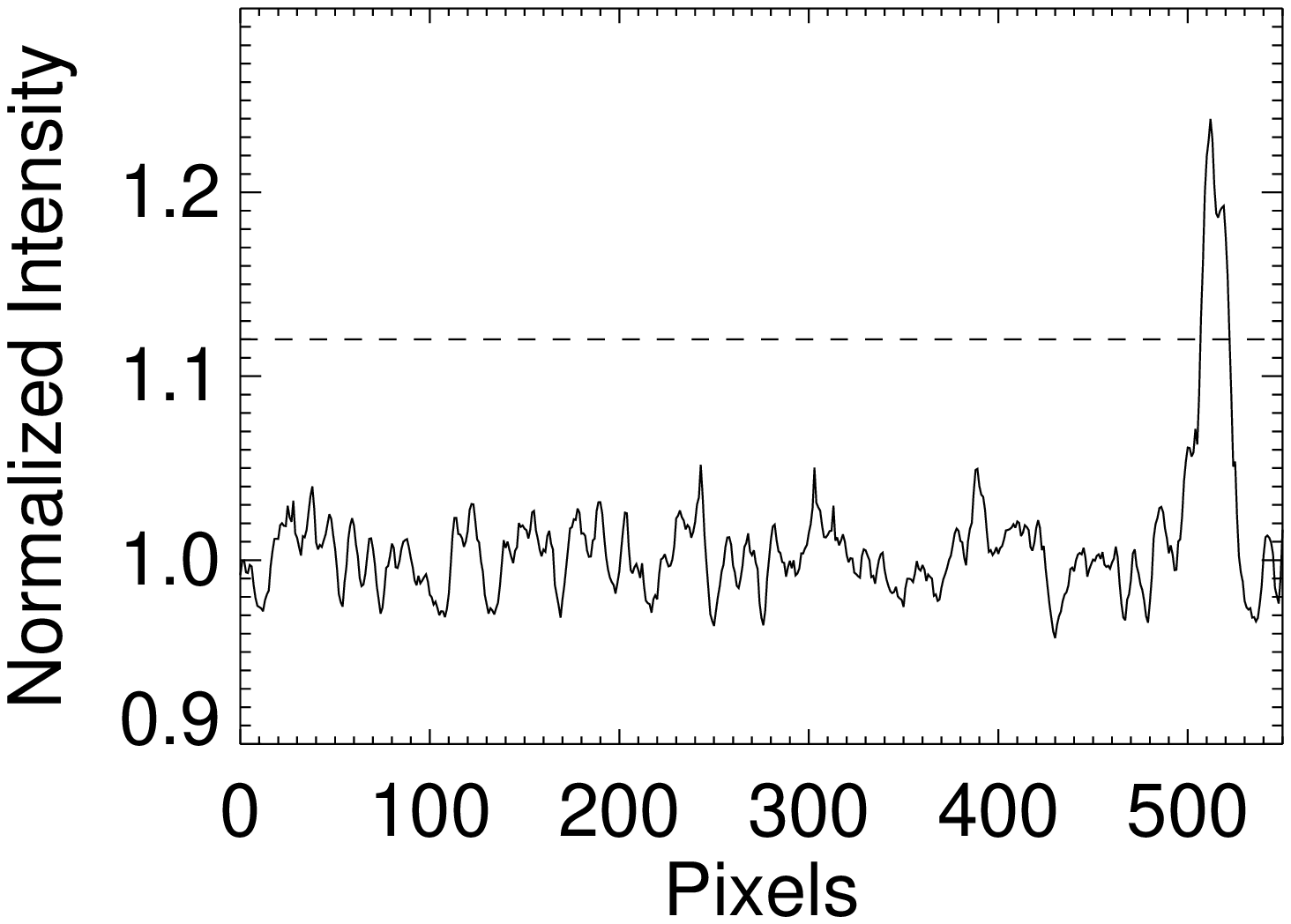}
\end{center}
\caption{Profile of row cuts (shown in Figure \ref{fig:3}(right) as two black lines) are shown here. 
Left: The plot is shown for the quiet region. The horizontal dashed and 
dash-dot-dot lines in the plot indicates the intensity contrast level of 
1.006 and 1.12 respectively. Right: The plot is shown for the row cut that includes 
the plage region. The dashed line is drawn to show the intensity contrast level of 1.12.}
\label{fig:4}
\end{figure*}

The background of the global fitted images have been normalized to a value equal to one.
Figure \ref{fig:4}(left) shows the intensity distribution along a row 
having only the network and intra-network elements extracted from the line indicated by 
a black in color in 
the middle of the image as shown in Figure \ref{fig:3}(right). The dashed line 
and the dash-dot-dot line in the plot represents
intensity contrast level of 1.006 and 1.12 respectively. The pixels with a spatial
scale of 2.5$\times$2.5 arc-sec$^{2}$ with an intensity contrast less than 1.006
represents background intra network. The locations of intensity contrast between
1.006-1.12 represents the network elements.
Figure \ref{fig:4}(right) shows intensity contrast distribution along a 
row, extracted from Ca-K image at a high latitude that has a small plage 
(can be seen as black line on the plage region in Figure \ref{fig:3}(right)).
The intensity contrast of the plage region is larger than 1.12.

\begin{figure*}
\begin{center}
\includegraphics[width=0.38\textwidth]{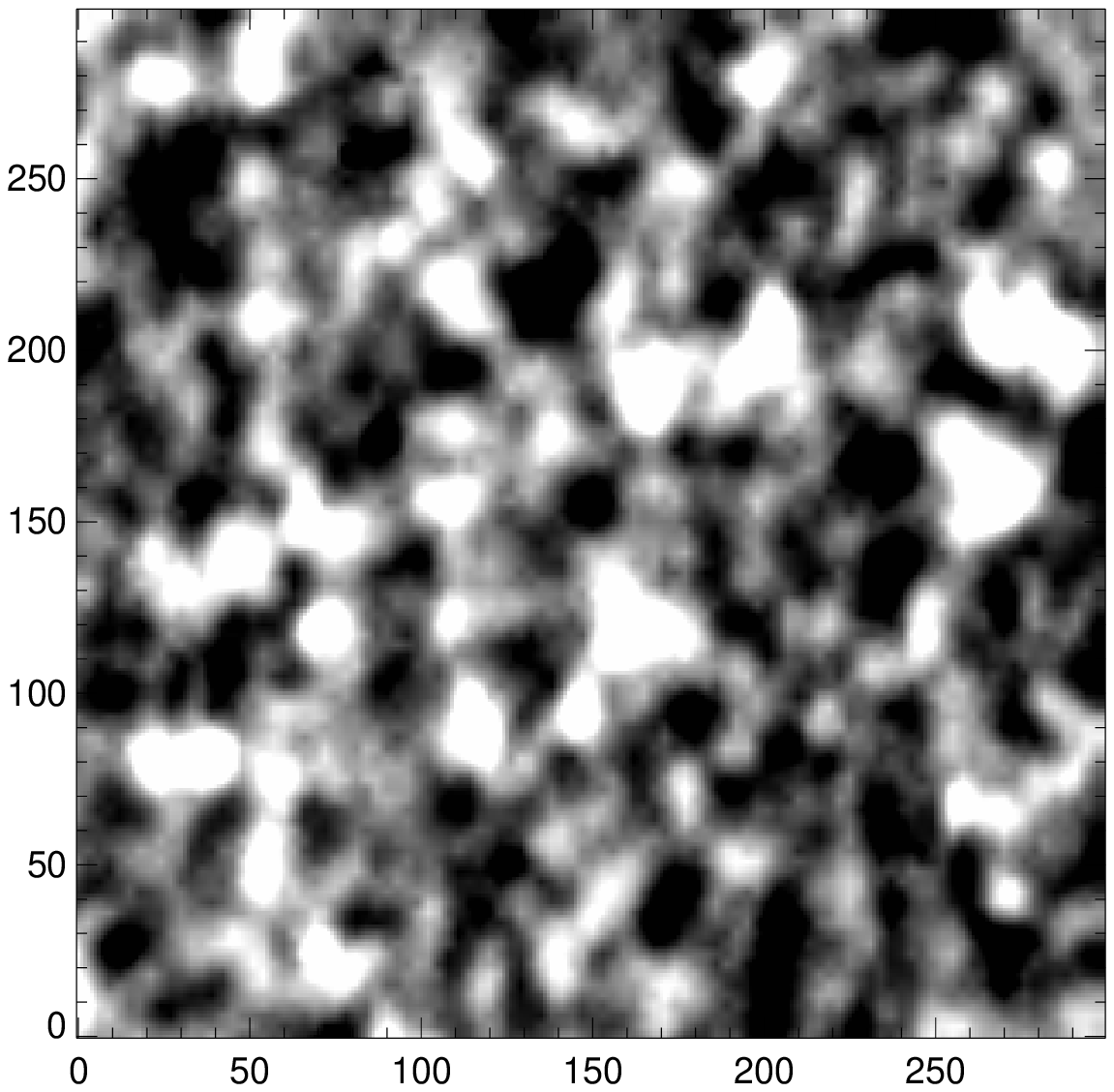} \includegraphics[width=0.38\textwidth]{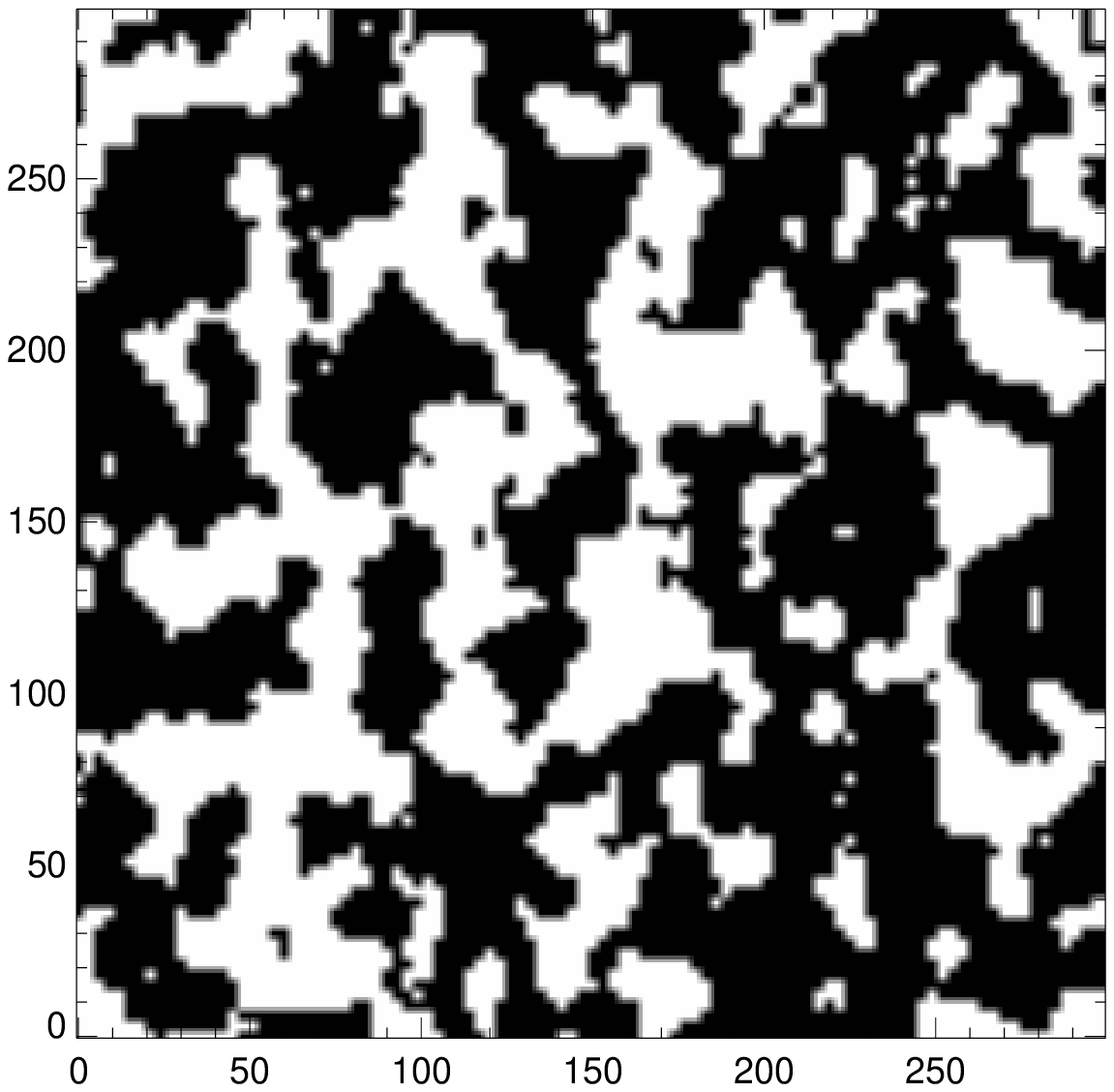}
\includegraphics[width=0.38\textwidth]{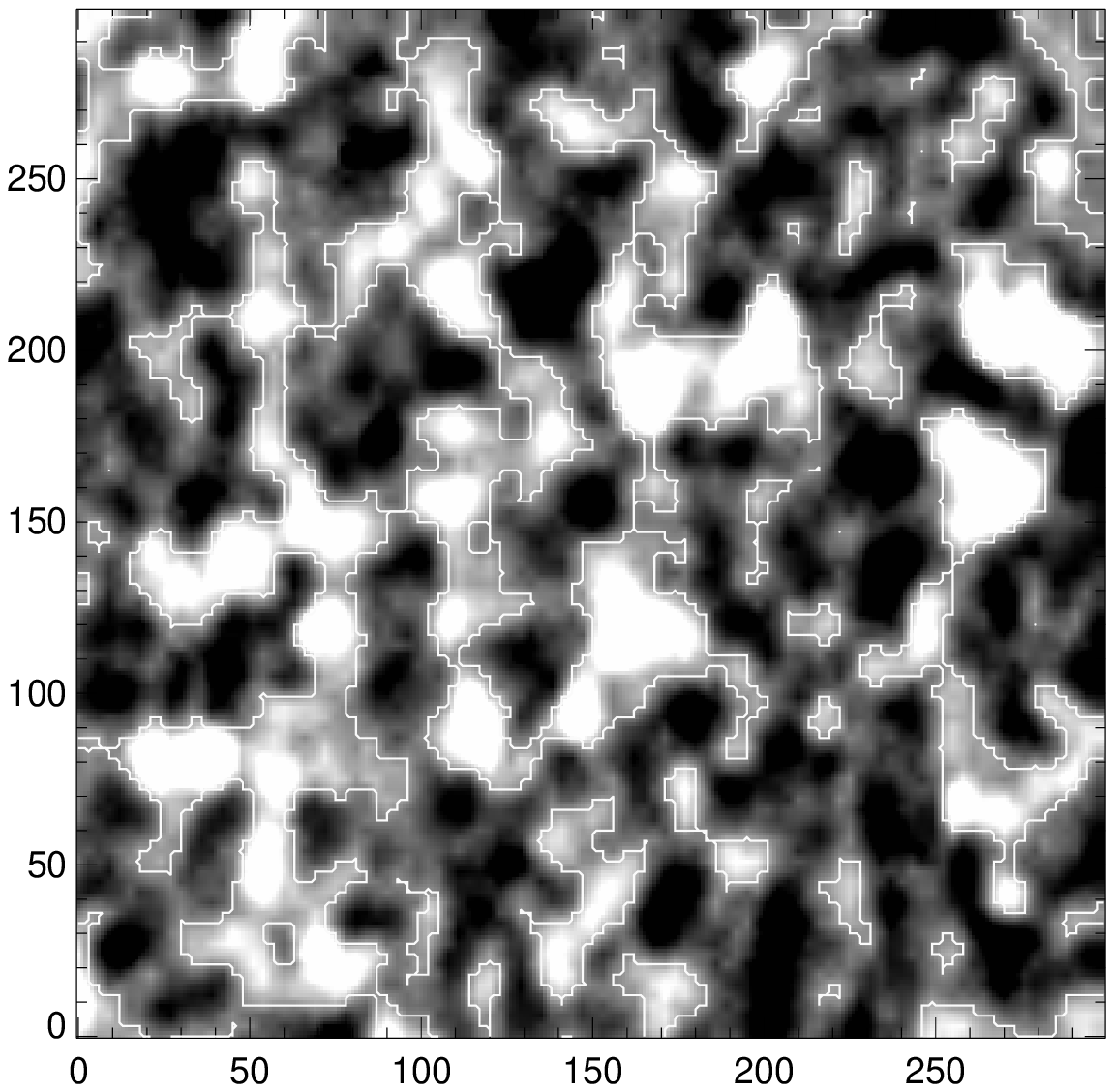} \\
\end{center}
\caption{Top left: A context image extracted from the center of the sun's image 
(shown as region I in Figure \ref{fig:3}(right)). 
Top right: A binary image of the left side image obtained by using the intensity 
contrast of 1.006 to 1.12. Bottom: Contours of
the top right image overlaid upon the top left image.} 
\label{fig:5}
\end{figure*}

\begin{figure*}
\begin{center}
\includegraphics[width=0.4\textwidth]{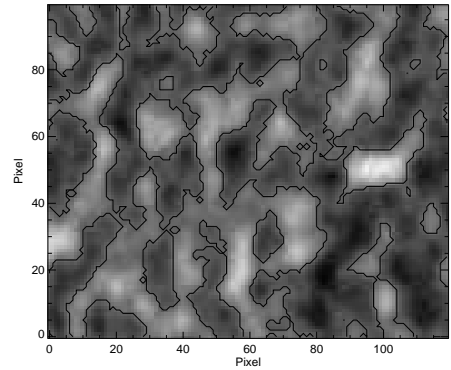}
\end{center}
\caption{Contours of the region obtained from the intensity contrast threshold values of 
1.006 (lower) and 1.12 (upper) overlaid upon the region II shown in Figure \ref{fig:3}(right).}
\label{fig:6}
\end{figure*}

\begin{figure*}
\begin{center}
\includegraphics[width=0.4\textwidth]{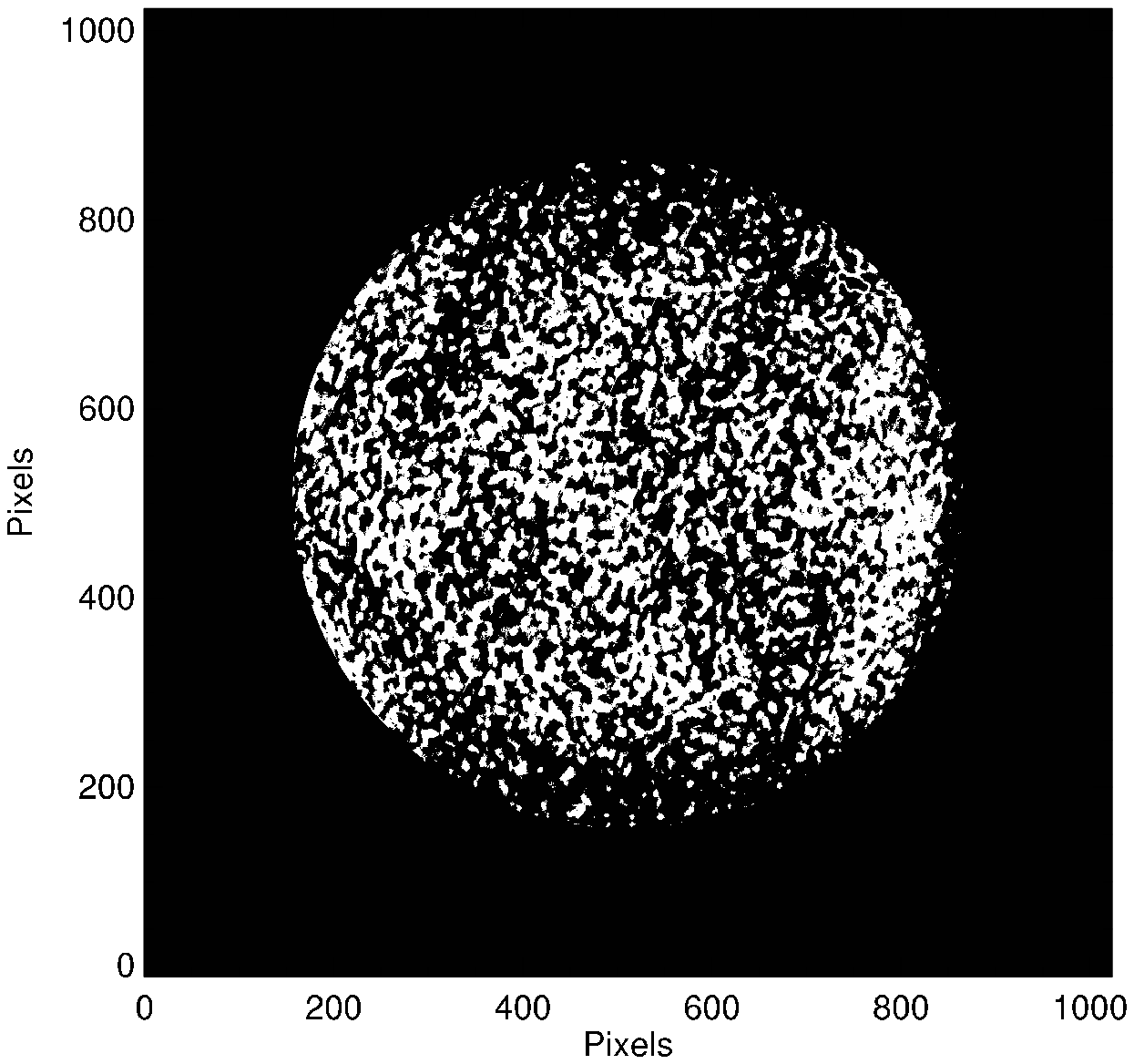}
\end{center}
\caption{Binary image for the extracted network elements using the upper and lower threshold values
for the whole disk.}
\label{fig:7}
\end{figure*}

In the following, we demonstrate that the selected threshold values of intensity contrast
are able to segregate the network elements and Ca-K
plage regions. Figure \ref{fig:5}(top left) shows the contest image extracted from the central 
portion of the solar image (box number I in Figure \ref{fig:3}(right)). The top right image
is the binary image of the left side image showing only the network elements extracted by
using a upper and lower threshold values 1.006 and 1.12 respectively. A comparison of the left
and right side images indicate that all the network elements
lie in this contrast range. The bottom image shows the contours
of the top right image overlaid upon the top left image. These maps show
that once the background has become uniform a fixed upper and
lower threshold are able to separate the network elements from others.
Figure \ref{fig:6} shows the contours of the 
detected network elements overlaid upon the region II shown in Figure \ref{fig:3}(right). This map
clearly shows that the contours coincides with the network
elements, suggesting that the method is able to identify the network
element features even near the limb. Figure \ref{fig:7} shows the extracted network 
elements on the whole disk. The extracted network pattern is uniform all over the disk
except some gaps are seen in the polar regions may be due to poor intensity contrast.

\begin{figure*}
\begin{center}
\includegraphics[width=0.4\textwidth]{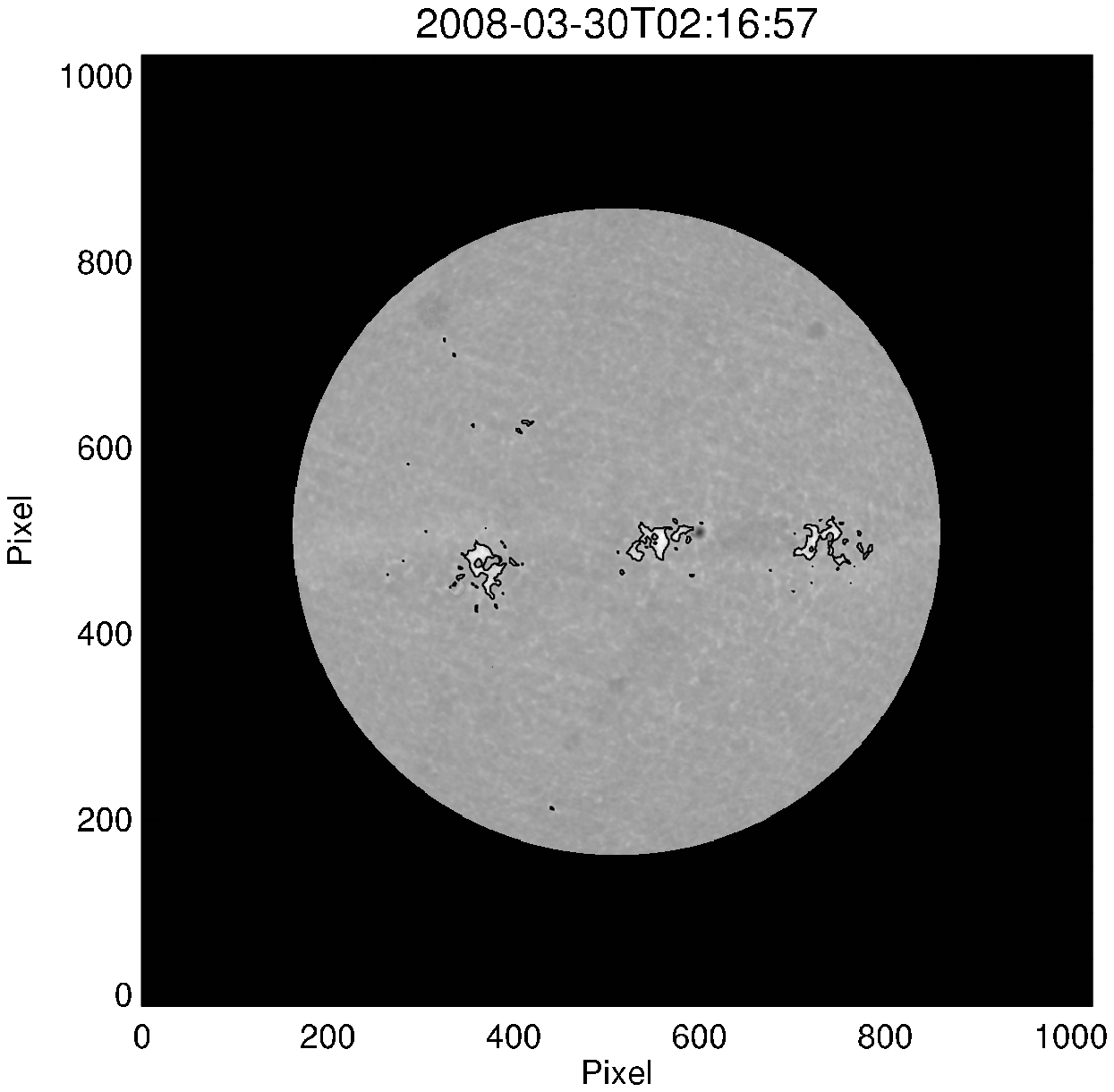}\includegraphics[width=0.3\textwidth]{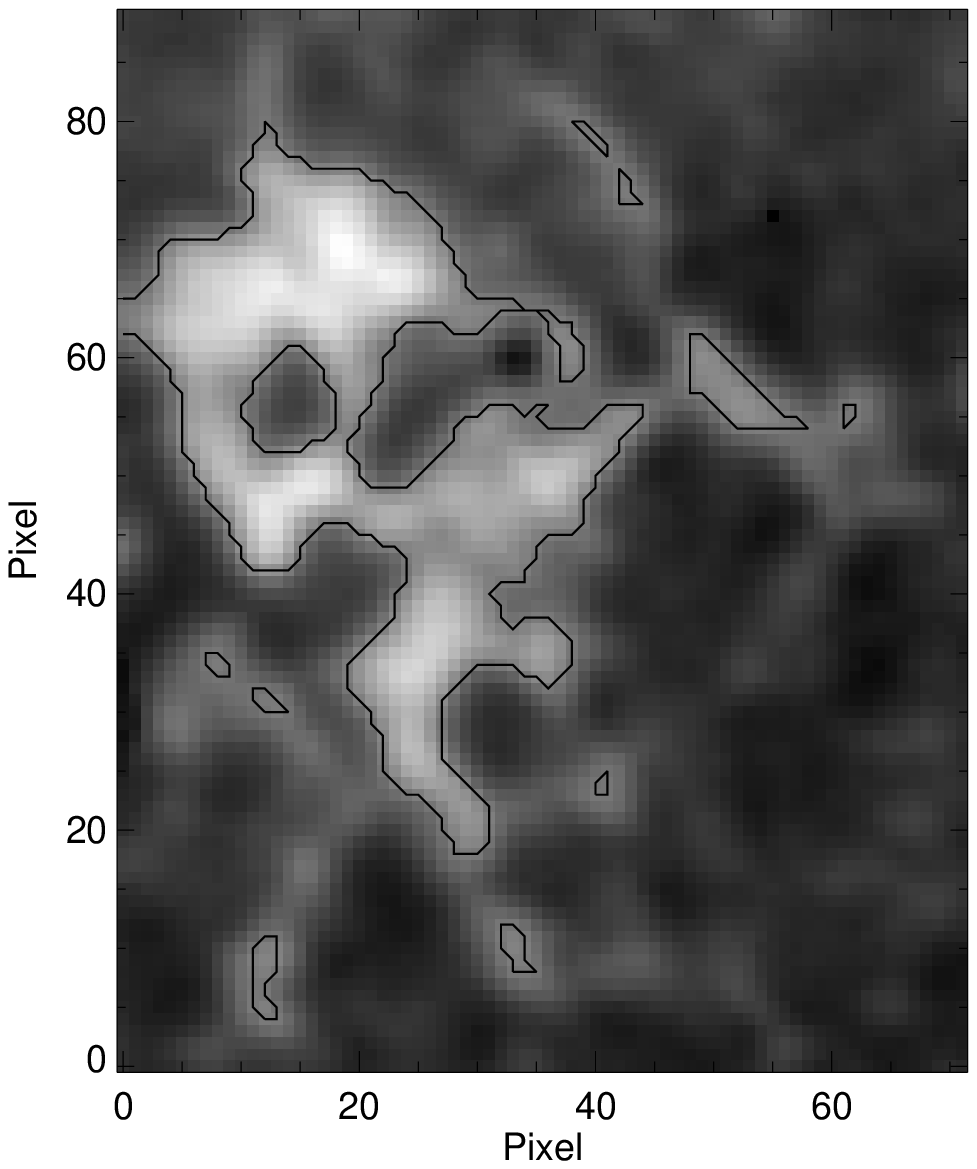}\\
\includegraphics[width=0.4\textwidth]{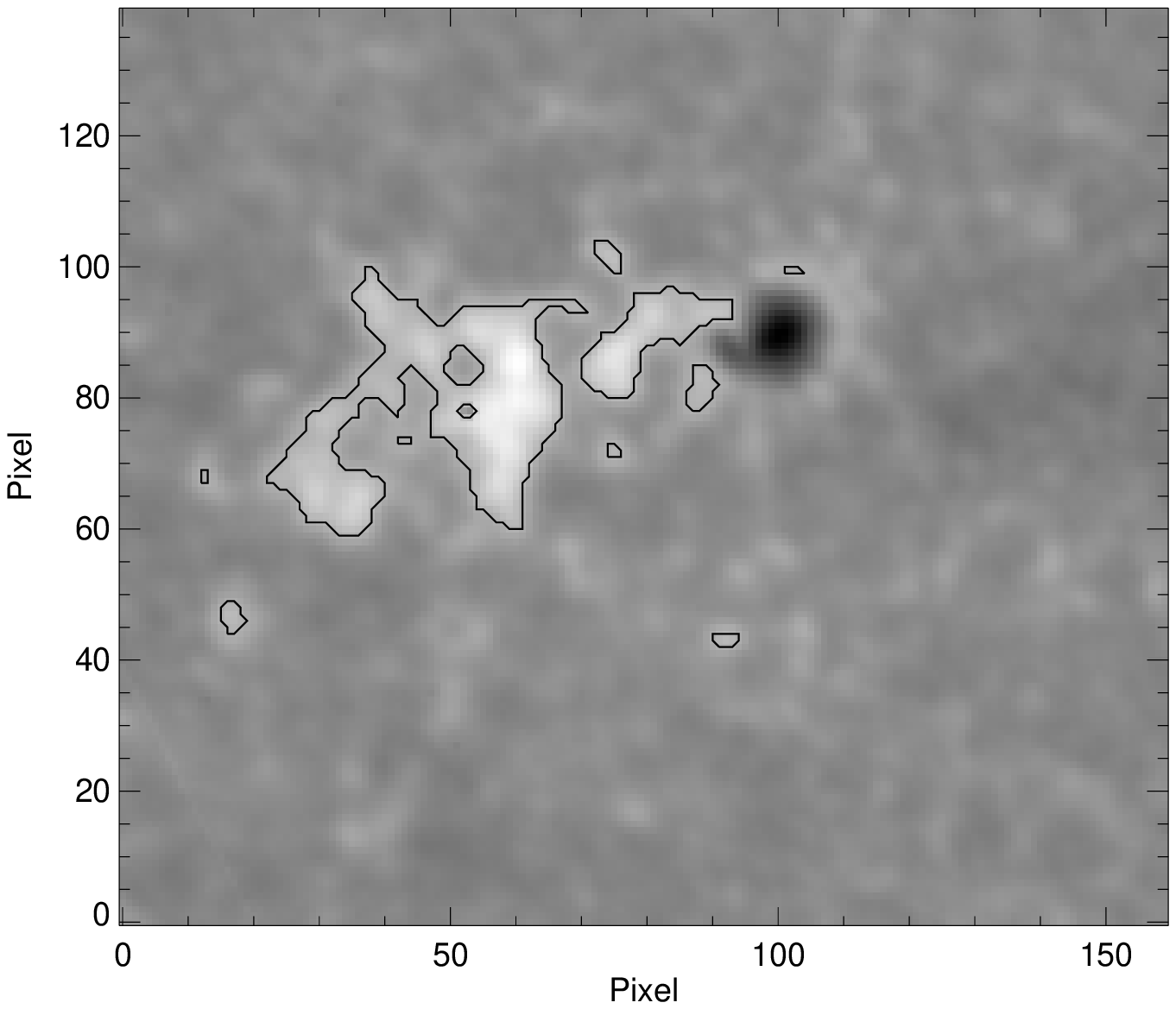}\includegraphics[width=0.4\textwidth]{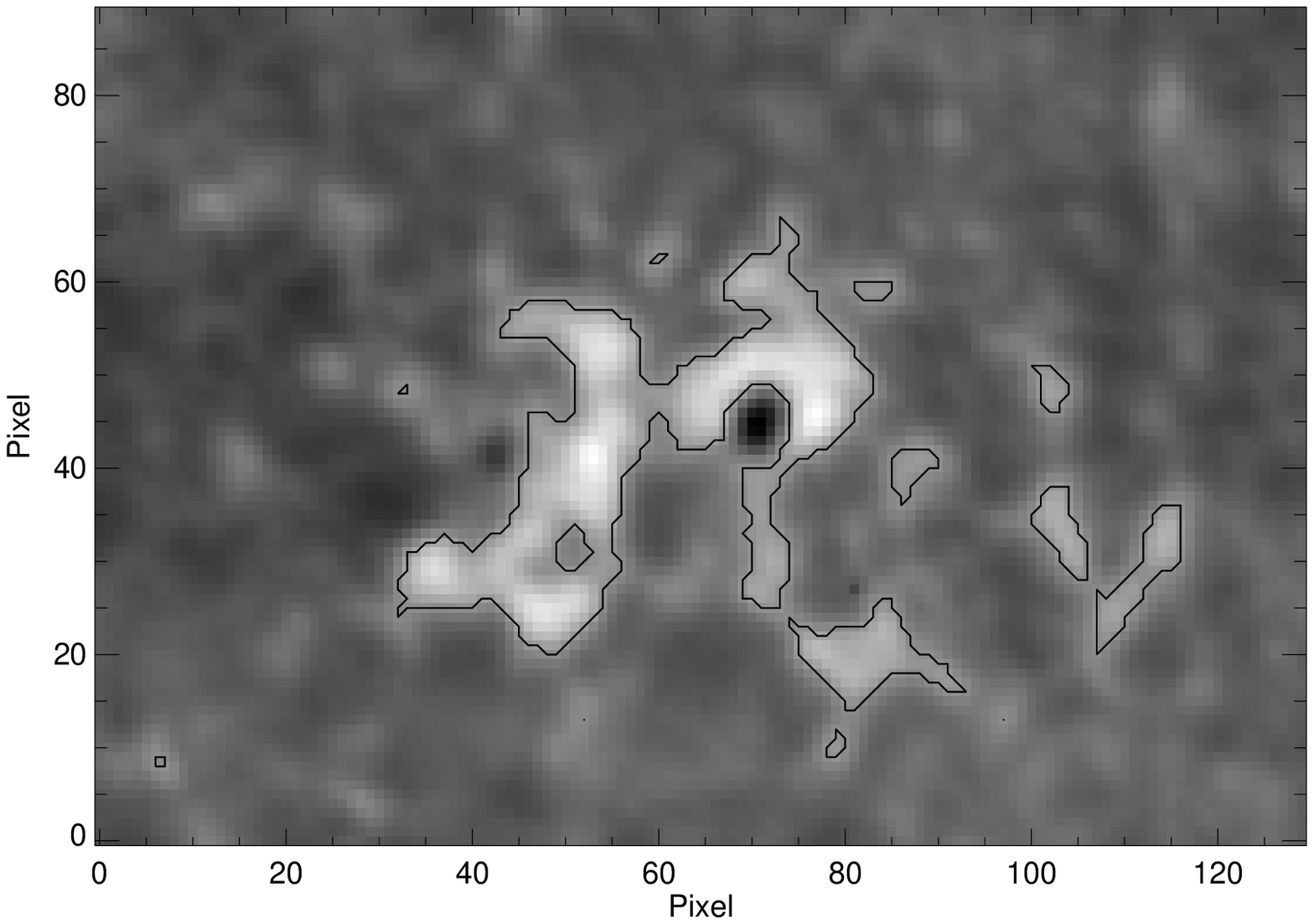}\\
\end{center}
\caption{Top left: Contours of plage region detected by the algorithm is overlaid upon the
full-disk image. The contours of the detcted plage regions overlaided upon the magnified portion
of the Ist (top-right), IInd (bottom left), IIIrd (bottom right) plage regions seen in full-disk image
from left to right.}
\label{fig:8}
\end{figure*}

Once the network elements have been identified by the algorithm based on the threshold
values of intensity contrast it is now simple task to extract the plage region. 
In the absence of enhanced and active network elements during this period we have identified 
plage regions with intensity contrast greater than 1.12. To make sure that this threshold 
value identifies 
only the plage region, the contours of the thresholded region obtained from
the algorithm is overlaided upon the sun's image and is shown in 
Figure \ref{fig:8}(top left).
Three different regions are magnified and the contours were overlaid upon it. This 
is shown in Figure \ref{fig:8}(top right, bottom left and right). The contours shown in 
Figure \ref{fig:8} demarcates the plages from rest of the regions.  
From these segregation of the individual regions it is now easy to estimate the 
network element area and plage area index.

\subsection{Network element area and plage area index}
The network element area index is expressed as a ratio of the area of the network elements
occupied on the sun to the area of the visible solar disk excluding the plage areas. 
We have excluded the plage area from the area of the visible solar disk while computing
network element index to study absolute variations of the index in this period by making the data 
uniform. We estimated the 
network element area index for 90\% of the solar disk. The rest 10\% close to the limb
is simply discarded as large projection effect and difficulty in extracting
network elements because of improper limb darkening correction at the edges.
We have computed the radius of the sun for a particular epoch. The 
area occupied by  all the pixels in contrast range from
1.006 to 1.12 was taken for network element area index calculation and the resulting network 
element area index is converted into percentage using the value of 90\% of visible area of 
the solar disk. 

\begin{figure*}
\begin{center}
\includegraphics[width=0.8\textwidth]{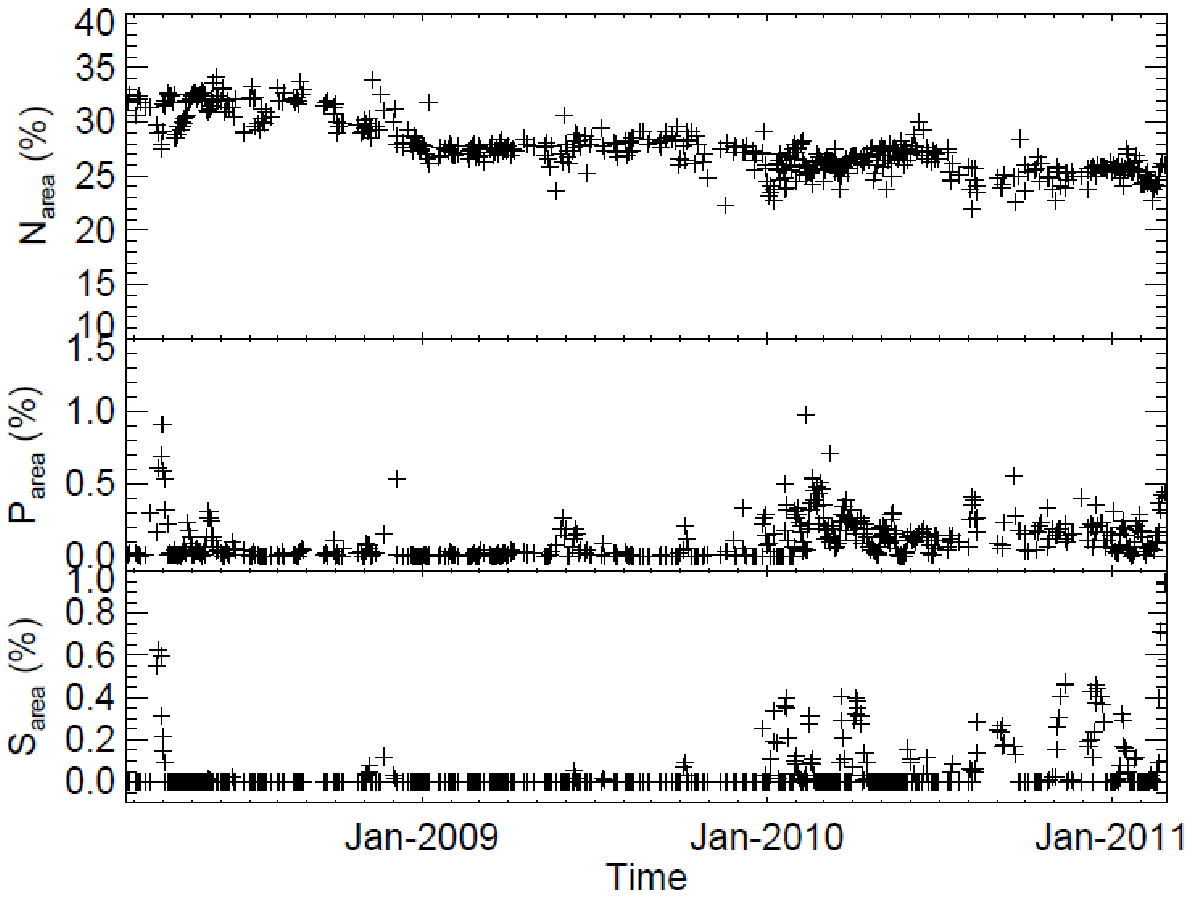} \\
\end{center}
\caption{Top: Daily average value of the network element area index (N$_{area}$). 
Middle: Plage area index (P$_{area}$) and Bottom: Sunspot area index (S$_{area}$)
expressed in terms of percentage are plotted over a period starting from February 2008 
to February 2011.}
\label{fig:9}
\end{figure*}
 
In order to study the variations in network element area index over the years we have plotted the 
daily average value of the network element area index for the period from Feb 2008 to Feb 2011 
of observations (Figure \ref{fig:9}).
The  plot shows that the network element occupies about 32\% of the solar disk from February 2008
till August 2008. Then after the network element area
decreases monotonically till February 2009. After that it remains flat with some small undulations 
in its value.  The error in measuring the network element area is 2\% of the daily
average network element area index. The error has been
estimated by varying the lower threshold value. The upper threshold value is kept at 1.12, as it 
is a large value and it did not affect our result larger than 0.02\%.
The network element area decreased by about 7\% starting from Feb 2008 to 2011.
The plage area during the solar minimum period and the ascending phase of solar cycle 24 occupies 
less than 1\% of the solar disk. It should be noted here that
the area occupied by the network element over the solar disk is not corrected for the projection
effect. The network elements covers the disk almost uniformly. We are
looking for the variations in the network element area index with time. While the value of
the network element area is affected by the projection effect equally over the observing period,
the percentage of the area may change marginally as the network element area 
changes proportionally to the visible area of the solar disk after the foreshortening correction. 
Thus, the observed differences in the 
area index over the 3 years (7\%) appear real and may not be due to foreshortening effect. 
We may consider this for the study of long term variations of network element area index in future.
The sunspot area index is also shown in the bottom of Figure \ref{fig:9} for comparison. Similar 
to the plage area index the sunspot area also covered less the 1\% of the solar disk during the 
extended minimum period.

\section{discussions}
By using Ca-K images obtained  from the Twin telescope at Kodaikanal and by applying a novel technique,
we were able to separate the network elements 
from the other structures such as back-ground and plage regions. After applying this technique, 
the contrast of the network elements is almost the same at all the disk positions and hence 
we were able to identify the network elements by using single upper and lower threshold values. From
the identified network elements we computed the network element area index over the
minimum period of solar cycle till the ascending phase of the solar cycle 24. 

It is not straight forward to compare our results with the results obtained by others 
in the past. This is mainly because of different pass bands of the filters used in the previous
observations, different spatial resolution of the instrument and seeing effects.  All these parameters 
affect the contrast of the network element features. Many of them have measured the network area 
for a few days and some of them have measured it for a few years and also included enhanced and 
active network where as our study deals with only quiet network elements. 
Singh, and Bappu~(\cite{Singh+Bappu1981}) reported that Ca-K network size is smaller in the quite region 
during the maximum phase of the solar cycle as compared to that during the minimum period.
Worden, White, and Woods~(\cite{Worden+White+Woods1998}) measured intensity contrasts of the
various features observed in Ca-K images and found that the intensity contrast of the plages,
enhanced network regions do not change over the solar cycle though their numbers will go down 
during solar minimum. They estimated that the plage and enhanced network cover the solar 
surface by 13\%  and 10\% respectively over the solar maximum. By using the Ca-K 
images from Rome PSPT, Ermolli, Berrilli, and Florio~(\cite{ermolli2003}) estimated that the 
network coverage over the disk
changes by 6\% over the ascending phase of the solar cycle 23, being small during the 
minimum and increased till solar maximum.  Walton, Preminger, and Chapman~(\cite{walton2003}) 
using San Fernando observatory
Ca-K images found that the area covered by the active region reduces by a factor of $\sim$ 20
from maximum to minimum. On the other hand the area covered by small regions (includes network)
reduces by a factor of about 2 from maximum to minimum of the activity cycle. 
Hagenaar, Schrijver and Title (\cite{hagenaar2003}) have observed a different magnetic 
flux concentrations in the network elements behaves differently with the solar cycle. 
Meunier (\cite{meunier2003}) reported that the magnetic flux in the network elements are
in phase with the solar cycle. Jin et~al. (\cite{jin2011}) obtained a 
result similar to Hagenaar, Schrijver and Title (\cite{hagenaar2003}) and they also reported
that the fractional area of quiet regions is anticorrelated with the solar cycle, but their total
magnetic flux is correlated with the solar cycle.
McIntosh et~al.~(\cite{mcintosh2011})
found that the mean supergranulation length scale changes from solar maximum to minimum.
They also have found that there is a difference of about 0.5~Mm in the mean radius of the
supergranulation during the current solar minimum compared to the previous solar minimum.
All these analysis mostly focus on the contrast of the network or the area of the network and
some of them use the magnetograms to measure the variations of the quiet sun flux with solar cycle.
In our study, we concentrated on the area occupied by the network elements over the solar 
disk during the minimum period and the ascending phase of solar cycle 24.  We 
found that about 7\% decrease in the disk coverage of the network element from minimum 
phase to ascending phase of solar cycle 24. The prolonged minimum period of solar cycle 23, 
there by a few or no active regions are observed on the sun. One of the main concept of the 
generation of the network fields is by the decay of the active regions. Since the number of 
active regions reduced to small or none, the magnetic fields of the network also could have 
reduced and hence the area occupied by the network elements. It could also be possible that 
we are observing the anticorrelated component of the quiet sun network here (Jin et~al. \cite{jin2011}). 
A long term network element data analysis is required to confirm this conjuncture.

\section{Summary}
\label{sec:summary}
Kodaikanal has a history of 100 years of data of Ca-K and white light images. To 
continue the availability of the data product for the next solar cycles we have developed 
a CCD based imaging system at Kodaikanal. The advantage of this system compared to the
previous system is that it can take a burst of images within a few hours. 
We have described a method 
to determine the network element area and plage area index separately. We applied this 
technique to the data obtained from the new telescope at Kodaikanal. Using this technique 
it was easy to separate the network and plage regions  
in the limb darkening corrected images. The result obtained 
from this technique indicate the network elements occupy about 30\% of the solar disk.
However, the network element area 
index slowly decreases by about 7\% during the period of about 3 years. The decrease in the daily 
average of network element area index over a period of 3~years could be related to the 
extended minima and continuing low activity phase.
We need a long time sequence of data to understand this variations
in the network element area index and plan to use digitized data for this study.
 
\subsection*{Acknowledgments}
We would like to thank the anonymous referee for his/her constructive comments on this paper.
Jagdev singh is thankful to Mr. F. Gabriel and his team for designing, fabricating 
and installing the telescope at Kodaikanal. We thank Mr. Anbazhagan and K. Ravi for 
developing the guiding system for the telescope. J. S. acknowledges the help of 
S. Muneer in the initial stages of the project and thanks to F. George, 
Mr. S. Ramamoorthy, P. Loganathan, P. Michael, P. Devendran, G. Hariharan, 
K. Fathima and S. Kamesh for their help to execute different parts of the project.

\label{lastpage}

\end{document}